\documentclass[nofootinbib,letterpaper]{revtex4}
\usepackage{graphicx}
\usepackage{dcolumn}
\usepackage{amsmath,amssymb,epsfig}
\usepackage{paralist}
\usepackage{comment}
\usepackage{graphicx}
\usepackage{multirow}
\usepackage{color,soul}

\allowdisplaybreaks

\renewcommand{\vec}[1]{\boldsymbol{\mathrm{#1}}}

\begin{document}


\title{Wave-theoretical description of the solar gravitational lens}

\author{Slava G. Turyshev}

\affiliation{\vskip 3pt
Jet Propulsion Laboratory, California Institute of Technology,\\
4800 Oak Grove Drive, Pasadena, CA 91109-0899, USA
}%

\date{\today}

\begin{abstract}

We discuss the propagation of electromagnetic (EM) waves in the post-Newtonian approximation of the general theory of relativity. We consider diffraction of EM waves in the static gravitational field of a massive monopole. We develop a wave-theoretical description of the solar gravitational lens (SGL) and show that with its enormous magnifying power of $\sim 10^{11}$ (for $\lambda=1~\mu$m) and angular resolution of $\lesssim 10^{-10}$~arcsec, the SGL may be used for direct megapixel imaging of an exoplanet.
\end{abstract}


\maketitle

Nature has presented us with a powerful ``instrument'' that we have yet to explore and learn how to use. The instrument is the solar gravitational lens (SGL), which takes advantage of the ability of the Sun's gravitational field to focus light  from faint, distant sources of significant scientific interest \cite{vonEshleman:1979}, such as a habitable exoplanet.  According to Einstein's general theory of relativity (GR), gravitation induces refractive properties on spacetime, causing a massive object to act as a lens by bending photon trajectories \cite{Einstein:1936}. To show this, we begin by considering the gravitational field of a static spherically symmetric distribution of matter in the post-Newtonian approximation of GR.  In the harmonic gauge \cite{Fock-book:1959}, $\partial_m (\sqrt{-g}g^{mn})=0$, the line element representing the static field of a gravitational monopole with a Schwarzschild radius of $r_g=2GM/c^2$ may be given in spherical coordinates $(r,\theta,\phi)$ as follows:
\begin{eqnarray}
ds^2&=&u^{-2}c^2dt^2-u^2\big(dr^2+r^2(d\theta^2+\sin^2\theta d\phi^2)\big), \qquad u=1+\frac{r_g}{2r}+{\cal O}(r_g^2,r^{-3}).
\label{eq:metric-mono}
\end{eqnarray}

To study light ray propagation in the metric (\ref{eq:metric-mono}), one usually \cite{Kopeikin-book-2011} takes the trajectory of a photon to be $\vec{x}(t)=\vec{x}_{0}+\vec{k}c(t-t_0)+\vec{x}_{\tt G}(t)+{\cal O}(G^2),$ where $\vec{x}_{0}$ is the initial position, $\vec{k}$ is the unperturbed wave vector, and $\vec{x}_{\tt G}(t)$ is the yet-unknown post-Newtonian term. We define the wave vector $K^m={dx^m}/{d\lambda}=K^0(1,{d\vec{x}}/{dx^0}),$ with $K^0={dx^0}/{d\lambda}$, where  $\lambda$  is the affine parameter along the ray's path. The wave vector obeys the geodesic equation: ${dK^m}/{d\lambda}+\Gamma^m_{kl}K^mK^l=0.$ Limiting discussion to the gravitational monopole, we obtain the solution for $\vec{x}_{\tt G}(t)$ and, thus, to the geodesic equation
{}
\begin{align}
\vec{x}(t)=\vec{b}_0+\vec{k}\ell-r_g\Big(\vec{k}\ln\frac{r+({\vec k}\cdot {\vec x})}{r_0+({\vec k}\cdot {\vec x}_0)}+\frac{\vec{b}_0}{b^2_0}\big(r+({\vec k}\cdot {\vec x})-r_0-({\vec k}\cdot {\vec x}_0)\big)\Big)+{\cal O}(r_g^2),
\label{eq:X-eq4**}
\end{align}
where $\ell=({\vec k}\cdot {\vec x}_{0})+c(t-t_0)$ and $\vec{b}_0=[\vec{k}\times[\vec{x}_0\times\vec{k}]]+{\cal O}(r_g)$ is the ray's impact parameter. 
Eq.~(\ref{eq:X-eq4**}) yields the deflection angle, $\delta$, which, for a distant source, $r=|{\vec x}|=\sqrt{b_0^2+\ell^2}\gg r_g$, is given by $\delta=|[{\vec k}\times({d{\vec x}}/{cdt})]|=2r_g/|{\vec b}_0| +{\cal O}(r_g^2).$

As a result, the gravitationally deflected rays of light passing from two sides of the lensing mass converge at a focus.
Of the solar system bodies, only the Sun is massive enough that the focus of its gravitational lens is within range of a realistic space mission. The effect is achromatic and, depending on the impact parameter, the SGL's focus is a semi-infinite line that begins at $\sim$547 astronomical units (AU) \cite{Turyshev-Andersson:2002}.  Eq.~(\ref{eq:X-eq4**}) describes the  trajectory of the light, but it tells nothing about its intensity.  Although this topic has been discussed earlier (see \cite{Turyshev-Andersson:2002} for review), the usual description is based on the geometric optics approximation, which, as expected, yields results that are divergent on the optical axis.  To investigate intensity changes due to the gravitational amplification of light, one needs to develop a wave-theoretical treatment of light propagation in gravity. In this paper, we provide such a description of the SGL.

Following  \cite{Landau-Lifshitz:1988},  in the spacetime (\ref{eq:metric-mono}), Maxwell's equations, $\partial_lF_{ik}+\partial_iF_{kl}+\partial_kF_{li}=0,$  $\partial_k\big(\sqrt{-g}F^{ik}\big)=-({4\pi}/{c}){\sqrt{-g}}j^i,$ describing the light propagation in a vacuum, reduce to the following set of equations for physical fields $(\vec D,\vec B)$:
{}
\begin{eqnarray}
[{\vec \nabla} \times {\vec D}]&=&- u^2\frac{1}{c}\frac{\partial \,{\vec B}}{\partial t}+{\cal O}(r_g^2), \quad {[}{\vec \nabla} \times{\vec B}]= u^2\frac{1}{c}\frac{\partial \,{\vec D}}{\partial t}+{\cal O}(r_g^2),
\quad ~{\vec \nabla} \cdot(u^2\,{\vec D})={\cal O}(r_g^2),
\quad
{\vec \nabla} \cdot(u^2{\vec B})={\cal O}(r_g^2),~~~
\label{eq:rotE_fl}
\end{eqnarray}
where ${\vec \nabla}$ is the ordinary differential operator with respect to flat space coordinates, and $\Delta=\nabla^2$. For a static metric, $g_{\alpha0}=0$, and the electromagnetic (EM) fields are related by $\vec  D=u\vec E$ and $\vec B=u\vec H$ (see Problem in \textsection 90 of \cite{Landau-Lifshitz:1988}).

Solution to Eqs.~(\ref{eq:rotE_fl}) may be given in the form of the Debye potentials \cite{Born-Wolf:1999}. In the case of a static monopole, the electric and magnetic Debye potentials, (${}^e{\hskip -1pt}\Pi,{}^m{\hskip -1pt}\Pi)$, are reduced to just one potential, $\Pi$, \cite{Herlt-Stephani:1975,Herlt-Stephani:1976}: $({}^e{\hskip -1pt}\Pi,{}^m{\hskip -1pt}\Pi)=\Pi(\cos\phi,\sin\phi)$. As a result, the solution for the vectors ${\vec D}={\vec D}(t,{\vec x})$ and ${\vec B}={\vec B}(t,{\vec x})$ may be given in the compact form:
{}
\begin{align}
  \left( \begin{aligned}
{D}_r& \\
{B}_r& \\
  \end{aligned} \right) =&  \left( \begin{aligned}
\cos\phi \\
\sin\phi  \\
  \end{aligned} \right) \,e^{-i\omega t}\alpha(r, \theta), &
    \left( \begin{aligned}
{D}_\theta& \\
{B}_\theta& \\
  \end{aligned} \right) =&  \left( \begin{aligned}
\cos\phi \\
\sin\phi  \\
  \end{aligned} \right) \,e^{-i\omega t}\beta(r, \theta), &
    \left( \begin{aligned}
{D}_\phi& \\
{B}_\phi& \\
  \end{aligned} \right) =&  \left( \begin{aligned}
\sin\phi \\
-\cos\phi  \\
  \end{aligned} \right) \,e^{-i\omega t}\gamma(r, \theta),
  \label{eq:DB-sol}
\end{align}
where the quantities $\alpha,\beta$ and $\gamma$ are given as below:
\begin{eqnarray}
\big\{\alpha; \beta; \gamma\big\}&=&\Big\{-\frac{1}{u^2r^2}\frac{\partial}{\partial \theta}\Big[\frac{1}{\sin\theta} \frac{\partial}{\partial\theta}\big[\sin\theta\,(r\,\Pi)\big]\Big];
~\frac{1}{u^2r}
\frac{\partial^2 \big(r\,{\hskip -1pt}\Pi\big)}{\partial r\partial \theta}+\frac{ik\big(r\,{\hskip -1pt}\Pi\big)}{r\sin\theta};
~-\frac{1}{u^2r\sin\theta}
\frac{\partial \big(r\,{\hskip -1pt}\Pi\big)}{\partial r}-\frac{ik}{r}
\frac{\partial\big(r\,{\hskip -1pt}\Pi\big)}{\partial \theta}\Big\},
\label{eq:albega}
\end{eqnarray}
and the Debye potential, $\Pi=\Pi(r,\theta)$, satisfies the following wave equation (denoting  $'=d/dr$):
\begin{eqnarray}
\Big(\Delta+\big(k^2u^4-u\big(\frac{1}{u}\big)''\big)\Big)\Big[\left( \begin{aligned}
\cos\phi \\
\sin\phi  \\
  \end{aligned} \right) \frac{\Pi}{u}\Big]=0 \quad {\rm or, ~equivalently, } \quad 
  \Big(\Delta+k^2\big(1+\frac{2r_g}{r}\big)+\frac{r_g}{r^3}\Big)\Big[\left( \begin{aligned}
\cos\phi \\
\sin\phi  \\
  \end{aligned} \right) \frac{\Pi}{u}\Big]={\cal O}(r_g^2).
\label{eq:Pi-eq+}
\end{eqnarray}

Eqs.~(\ref{eq:DB-sol})--(\ref{eq:Pi-eq+}) provide a complete solution for the diffraction problem. If the incident wave is known, by using the approach developed in the Mie theory \cite{Born-Wolf:1999}, one can find the scattered wave. To find the solution for the incident wave, we use (\ref{eq:rotE_fl}) to see that, for instance, ${\vec D}$ has to obey the following wave equation (the equation for ${\vec B}$  has an identical form):
{}
\begin{eqnarray}
\Delta {\vec D} &=&u^4\frac{\partial^2 {\vec D}}{c^2\partial t^2}+[[{\vec\nabla}\times{\vec D}]\times {{\vec \nabla}\ln u^2}]- {\vec \nabla} \big({\vec D}\cdot {\vec \nabla}\ln u^2\big)+{\cal O}(r_g^2).
\label{eq:wave-eq_uE0}
\end{eqnarray}
We look for a solution to (\ref{eq:wave-eq_uE0}) in the  form ${\vec D}=\psi\vec d e^{-i\omega t} $, where $\psi\equiv \psi(\vec r)$ is some scalar function, $\vec d\equiv {\vec d}(\vec r)$ is a unit vector specifying the direction of the wave's propagation and its polarization, and $\omega$ is the frequency of the wave. One can see that, for a compact source of a static weak gravitational field described by (\ref{eq:metric-mono}) and for the case of propagation of the high-frequency EM waves (i.e., $k=\omega/c\rightarrow\infty$), Eq.~(\ref{eq:wave-eq_uE0}) yields two equations:
\begin{eqnarray}
\Delta \psi+k^2\Big(1+\frac{2r_g}{r}\Big)\psi&=&{\cal O}(r_g^2),
\qquad
({\vec \nabla}\psi\cdot{\vec \nabla})\,{\vec d}=\frac{r_g}{r^3}\Big\{({\vec d}\cdot{\vec x})\,{\vec \nabla}\psi-{\textstyle\frac{1}{2}}({\vec \nabla}\psi\cdot{\vec x}){\vec d}\Big\}+{\cal O}(r_g^2).
\label{eq:wave-eik2}
\end{eqnarray}
Eqs.~(\ref{eq:wave-eik2}) provide a complete description of an EM wave propagating in weak, static gravity. The left of Eq.~(\ref{eq:wave-eik2}) determines the change in the intensity of the EM radiation, while the one on the right determines the change in the direction of the wave propagation and describes the polarization changes of the EM wave along the path.

To establish the wave properties of light, we need to go beyond (\ref{eq:X-eq4**}), which describes light as a massless particle traveling along a geodesic, and solve equations (\ref{eq:wave-eik2}). Thus, the problem of image formation by the SGL amounts to solving (\ref{eq:wave-eik2}) for astronomically relevant conditions. We begin with the left of Eq.~(\ref{eq:wave-eik2}), which is formally similar to the time-independent Schr\"odinger equation \cite{Messiah:1968}.
This equation has a solution regular at the origin in the form of
{}
\begin{eqnarray}
\psi(\vec r)=\psi_0e^{ikz}{}_1F_1\big(ikr_g, 1, ik(r-z)\big),
\label{eq:psi_hyp_geom+}
\end{eqnarray}
where ${}_1F_1(\alpha,\beta,z)$ is the confluent hypergeometric function of the first kind, $\psi_0=e^{\frac{\pi}{2}kr_g}{\Gamma(1-ikr_g)}$ is the  normalization constant, such that $\psi^2\rightarrow1$, while $|k(r-z)|\rightarrow\infty$. This solution is for a wave coming from a large distance along the $z$ axis. It generalizes the plane wave solution $\psi_0(\vec r)=e^{ikz}$ that is used to describe EM wave propagation in Euclidean spacetime. All the important corrections to $\psi$ due to weak gravity are contained in the ${}_1F_1$ function.
For large distances from the deflector, Eq.~(\ref{eq:psi_hyp_geom+}) has the following asymptotic form (see \cite{Messiah:1968} and with the help of  \cite{Abramovitz-Stegan:1965}):
{}
\begin{eqnarray}
\psi= e^{ik(z-r_g\ln k(r-z))}+\frac{r_g}{r-z}\frac{\Gamma(1-ikr_g)}{\Gamma(1+ikr_g)}e^{ik\big(r+r_g\ln k(r-z)\big)}+{\cal O}(r_g^2),
\label{eq:sol-hgeom*=/*=}
\end{eqnarray}
where the first term represents an incident Coulomb-modified wave \cite{Messiah:1968}, while the second  term is the scattered wave.

Given the solution for the amplitude of the incident wave from (\ref{eq:sol-hgeom*=/*=}), we can proceed to solve the second of (\ref{eq:wave-eik2}). 
The parameter $\ell$ introduced in (\ref{eq:X-eq4**}) along the unperturbed direction of the ray's path allows to represent this equation  as
{}
\begin{eqnarray}
\frac{d \,{\vec d}}{d\ell}&=&\frac{r_g}{r^3}\Big\{({\vec d}\cdot{\vec x})\,{\vec k}-{\textstyle\frac{1}{2}}({\vec k}\cdot{\vec x})\,{\vec d}\Big\}+{\cal O}(r_g^2).
\label{eq:wave-eik2*!}
\end{eqnarray}
Similarly to (\ref{eq:X-eq4**}), we write $\vec d=d_{||}\,{\vec k}+{\vec d}_{\perp0}+{\vec d}_{\tt G}+{\cal O}(r_g^2),$ where $d_{||0}=({\vec d}\cdot {\vec k})+{\cal O}(r_g)$ and ${\vec d}_{\perp0}=[{\vec k}\times[{\vec d}\times{\vec k}]]+{\cal O}(r_g)$ and ${\vec d}_{\tt G}$ is the post-Newtonian part of vector ${\vec d}$.  As a result, Eq.~(\ref{eq:wave-eik2*!}) takes the form
{}
\begin{eqnarray}
\frac{d \,{\vec d}_{\tt G}}{d\ell}&=&\frac{r_g}{(b_0^2+\ell^2)^{3/2}}\Big\{\Big({\textstyle\frac{1}{2}}d_{||0}\ell+({\vec d_{\perp0}}\cdot{\vec b}_0)\Big)\,{\vec k}-{\textstyle\frac{1}{2}}\ell\,{\vec d}_{\perp0}\Big\}+{\cal O}(r_g^2).
\label{eq:wave-eik2**!}
\end{eqnarray}
We introduce a heliocentric Cartesian coordinate system $(x,y,z)$  with unit vectors $({\vec e}_x, {\vec e}_y, {\vec e}_z)$.  We take the $z$ axis to be directed along the vector ${\vec k}$, while the $x$ and $y$ axes directed along the vectors ${\vec e}=[[{\vec k}\times{\vec n}]\times{\vec k}]$ and ${\vec p}=[{\vec k}\times{\vec n}]$, correspondingly, where ${\vec n}={\vec x}/r$; in other words $({\vec e}_x, {\vec e}_y, {\vec e}_z)\equiv({\vec e}, {\vec p}, {\vec k})$.  In this coordinate system, $d_{||0}=d_{z0}$, ${\vec d}_{\perp0}=(d_{x0},d_{y0},0)$, ${\vec b}_{0}=[{\vec k}\times[{\vec x}\times{\vec k}]]=(x,y,0)+{\cal O}(r_g)$, and, thus, $({\vec d}_{\perp 0}\cdot {\vec b}_0)=d_{x0}x+d_{y0}y+{\cal O}(r_g)$. We choose the components of the incident wave so that it represents a transverse-electric wave requiring: $d_{z0}=d_{y0}=0$ and $d_{x0}=1$. We determine the components of the incident $\vec D$ field in the heliocentric spherical coordinate system $(r,\theta,\phi)$,
{}
\begin{eqnarray}
\big\{D^{\tt inc}_{r}; D^{\tt inc}_\theta; D^{\tt inc}_\phi \big\}& =&
\Big\{-\,\frac{\cos\phi}{iukr}\frac{\partial \psi_i}{\partial\theta}; ~u^{-1}\cos\phi \big(\cos\theta-\frac{r_g}{r}\big)\psi_i; ~
- u\sin\phi  \psi_i\Big\}e^{-i\omega t} +{\cal O}(r_g^2),
  \label{eq:vec_D_sol+}
\end{eqnarray}
where $\psi_i=e^{ik(r\cos\theta-r_g\ln kr(1-\cos\theta))}$ is the incident wave from (\ref{eq:sol-hgeom*=/*=}). We can obtain a similar solution for ${\vec B}^{\tt inc}$.

We now need to find the EM field, which for $r \rightarrow \infty, \theta \sim \pi$ has the same asymptotic behavior as the incident field (\ref{eq:vec_D_sol+}), but which is regular everywhere, for all values of $\theta$ and $r$. As the wave function (\ref{eq:psi_hyp_geom+}) gives the correct asymptotic expression at small angles, the required field may be constructed using (\ref{eq:psi_hyp_geom+}). To determine $\Pi$, we use the expressions for the incident wave (\ref{eq:vec_D_sol+})  and relate them to (\ref{eq:DB-sol}). One of Eqs.~(\ref{eq:DB-sol}) needs to be solved, e.g., $D_{r}$.  To find the solution in all regions we  extend (\ref{eq:vec_D_sol+}) by taking, instead of $\psi_i$, the entire solution for $\psi$ from   (\ref{eq:psi_hyp_geom+}). The exact solution (\ref{eq:psi_hyp_geom+})  should differ from the incident wave (\ref{eq:vec_D_sol+}) only for the outgoing waves. The amplitudes of incident waves should be equal.  Equations (\ref{eq:vec_D_sol+}) indicate that $D_{r}=-e^{-i\omega t}\,\frac{\cos\phi}{iukr}\frac{\partial \psi}{\partial\theta}$
is a suitable definition of the wanted regular field.  From (\ref{eq:DB-sol}), (\ref{eq:vec_D_sol+}), this yields
{}
\begin{eqnarray}
D_{r}&=&-e^{-i\omega t}\,\frac{\cos \phi}{u^2r^2}\frac{\partial}{\partial \theta}\Big[\frac{1}{\sin\theta} \frac{\partial}{\partial\theta}\big[\sin\theta\,(r\,\Pi)\big]\Big]=-e^{-i\omega t}\,\frac{\cos\phi}{iukr}\frac{\partial \psi}{\partial\theta},
\label{eq:vec_D_r*}
\end{eqnarray}
where $\psi$ has the form (\ref{eq:psi_hyp_geom+}).
As a result, (\ref{eq:vec_D_r*}) yields an equation to determine the Debye potential $\Pi$,
{}
\begin{eqnarray}
\frac{\partial}{\partial \theta}\Big[\frac{1}{\sin\theta} \frac{\partial}{\partial\theta}\big[\sin\theta\,\Pi\big]\Big]=-\frac{iu}{k}\frac{\partial \psi}{\partial\theta}+{\cal O}(r_g^2).
\label{eq:vec_D_r*+}
\end{eqnarray}
After integrating this equation with respect to $\theta$, we obtain the solution for the Debye potential as
{}
\begin{equation}
\Pi(\vec r)=-\psi_0\frac{iu}{k}\frac{1-\cos\theta}{\sin\theta}e^{ikz}\Big({}_1F_1[1+ikr_g,2,ikr(1-\cos\theta)]-{}_1F_1[1+ikr_g,2,2ikr]\Big)+{\cal O}(r_g^2),
\label{eq:sol-Pi0}
\end{equation}
which gives the Debye potential in terms of the Coulomb wave function $\psi$ from (\ref{eq:psi_hyp_geom+}), i.e., essentially in terms of the confluent hypergeometric series. It can be shown that the second term in (\ref{eq:sol-Pi0}) is negligible. The EM field and the Poynting vector due to this term are orders of magnitude (factor $1/\sqrt{kr_g}$) smaller than those originating from the first term. This term makes it possible to avoid singular behavior of $\Pi$ at the axis $\theta=\pi$ and is important only near this axis \cite{Herlt-Stephani:1976}.

Using the solution for $\Pi$ from (\ref{eq:sol-Pi0}), one can now compute all the quantities in (\ref{eq:albega}). To discuss the relevant results, it is convenient to introduce another, cylindrical coordinate system $(\rho,\phi,z)$. In the far field, $r \gg r_g$, we do this by introducing $\rho=R\sin\theta,   \phi=\phi,  z=R\cos\theta,$ with $R=\sqrt{\rho^2+z^2},  \theta=\arctan({\rho}/{z})$, and the corresponding line element
{}
\begin{eqnarray}
ds^2&=&u^{-2}c^2dt^2-\big(d\rho^2+\rho^2d\phi^2+u^2dz^2\big)+{\cal O}(r_g^2).
\label{eq:cyl_coord}
\end{eqnarray}
In this coordinate system,  the components of the EM field are:
{}
\begin{align}
  \left( \begin{aligned}
{D}_\rho& \\
{B}_\rho& \\
  \end{aligned} \right) =&  \left( \begin{aligned}
\cos\phi \\
\sin\phi  \\
  \end{aligned} \right) \,e^{-i\omega t}a(r, \theta), &
    \left( \begin{aligned}
{D}_z& \\
{B}_z& \\
  \end{aligned} \right) =&  \left( \begin{aligned}
\cos\phi \\
\sin\phi  \\
  \end{aligned} \right) \,e^{-i\omega t}b(r, \theta), &
    \left( \begin{aligned}
{D}_\phi& \\
{B}_\phi& \\
  \end{aligned} \right) =&  \left( \begin{aligned}
\sin\phi \\
-\cos\phi  \\
  \end{aligned} \right) \,e^{-i\omega t}\gamma(r, \theta),
  \label{eq:DB-sol-cyl}
\end{align}
with
$a(r,\theta)=u^{-1}\sin\theta \,\alpha(r,\theta)+\cos\theta \,\beta(r,\theta)$,  $b(r,\theta)=\cos\theta \,\alpha(r,\theta)-u\sin\theta\,\beta(r,\theta).$
Using (\ref{eq:albega}) and (\ref{eq:sol-Pi0}) and referring to \cite{Abramovitz-Stegan:1965} for the properties of the confluent hypergeometric functions, the solutions for functions $a, b, \gamma$ take the form:
{}
\begin{eqnarray}
a(r, \theta)&=&\frac{1}{u}\psi_0e^{ikz}\Big\{F[1]\Big(1-\frac{r_g}{2r}\sin^2\theta\Big)+F[2]\,\Big(\frac{1-\cos\theta}{\sin^2\theta}\cos\theta\big(1-\cos\theta +\frac{r_g}{r}\big)-
ikr_g\big(1-\cos\theta\big)\Big\}+{\cal O}(r_g^2),
\label{eq:a0}\\
b(r, \theta)&=&-\frac{1}{u}\psi_0e^{ikz}\sin\theta\Big\{F[1]\frac{r_g}{2r}\cos\theta+F[2]\,\Big(\frac{1-\cos\theta}{\sin^2\theta}u\big(1-\cos\theta +\frac{r_g}{r}\big)+
ikr_g\Big)\Big\}+{\cal O}(r_g^2),
\label{eq:b0}\\[0pt]
\gamma(r, \theta)&=&-u\psi_0e^{ikz}\Big\{ F[1]+
 F[2]\,\frac{1-\cos\theta}{\sin^2\theta}\Big(1-\cos\theta -\frac{r_g}{r}\Big)\Big\}+{\cal O}(r_g^2),
\label{eq:gamma1*}
\end{eqnarray}
where we defined $F[1]\equiv{}_1F_1[ikr_g,1,ikr(1-\cos\theta)]$ and $F[2]\equiv{}_1F_1[1+ikr_g,2,ikr(1-\cos\theta)].$

The components of the Poynting vector,
${\vec S}=[{\vec E}\times{\vec H}]= u^{-2}[\operatorname{Re}({\vec D})\times \operatorname{Re}({\vec B})]$, in cylindrical coordinates are
{}
\begin{eqnarray}
S_\rho&=&u^{-2}{\rm Re}(e^{-i\omega t}\gamma) \, {\rm Re}(e^{-i\omega t}b),\qquad
S_z=-u^{-2}{\rm Re}(e^{-i\omega t}\gamma) \, {\rm Re}(e^{-i\omega t}a),
\qquad
S_\phi=0.
\label{eq:S_ph}
\end{eqnarray}
Using  Eqs.~(\ref{eq:a0})--(\ref{eq:gamma1*}), after time averaging, we get the Poynting vector (\ref{eq:S_ph})  for high frequencies, $kr\rightarrow\infty$, as
{}
\begin{eqnarray}
{\bar S}_\rho&=&u^{-2}\frac{1}{2}\psi_0^2\sin\theta\Big\{F[1]F^*[1]\frac{r_g}{2r}\cos\theta+F[2]F^*[2]\Big(\frac{1-\cos\theta}{\sin^2\theta}\Big)^2u\big(1-\cos\theta\big)^2+\nonumber\\
&+&
{\textstyle\frac{1}{2}}\Big(F[1]F^*[2]+F^*[1]F[2]\Big)\frac{1-\cos\theta}{\sin^2\theta}\Big(1-\cos\theta+\frac{r_g}{2r}\sin^2\theta\Big)-
{\textstyle\frac{1}{2}}i\Big(F[1]F^*[2]-F^*[1]F[2]\Big)kr_g
\Big\},
\label{eq:S_rh*3}\\
{\bar S}_z&=&u^{-2}\frac{1}{2}\psi_0^2\Big\{F[1]F^*[1]\big(1-\frac{r_g}{2r}\sin^2\theta\big)+F[2]F^*[2]\Big(\frac{1-\cos\theta}{\sin^2\theta}\Big)^2\big(1-\cos\theta\big)^2\cos\theta+\nonumber\\
&+&
{\textstyle\frac{1}{2}}\Big(F[1]F^*[2]+F^*[1]F[2]\Big)\big(1-\frac{r_g}{2r}(1-\cos\theta)\big)\big(1-\cos\theta\big)+
{\textstyle\frac{1}{2}}i\Big(F[1]F^*[2]-F^*[1]F[2]\Big)kr_g(1-\cos\theta)
\Big\},~~~
\label{eq:S_z*3}
\end{eqnarray}
where ${\bar S}_z, {\bar S}_\rho,$ and ${\bar S}_\phi=0$ are the time-averaged components of the Poynting vector in the coordinate system (\ref{eq:cyl_coord}) and are all accurate to ${\cal O}(r_g^2,(kr)^{-1})$; also,  $F^*$ denotes complex conjugate of $F$.

All properties of the diffraction field are contained in the formulas (\ref{eq:S_rh*3})--(\ref{eq:S_z*3}), covering all distances and angles around the Sun. Extracting these properties is somewhat complicated, because too many parameters enter into the structure of this interference pattern: the radial distance $r \sim z$, the distance $\rho = r\sin\theta$ from the axis $\theta = 0$ in the image plane, the frequency $\omega$, and, if we ask for the visual image of an exoplanet, the aperture of our telescope. We therefore confine ourselves to the most interesting results in the interference region in the  vicinity of the optical axis.

Consider  the hypergeometric series $F[1]$  at small angles $\theta\approx \rho/z\ll1$: for $kr\gg1$,  from \cite{Abramovitz-Stegan:1965} we have  $F[1]=
J_0(2\sqrt{x})+{\cal O}((kr_g)^{-1})$ and $F[2]=(1/\sqrt{x})J_1(2\sqrt{x})+{\cal O}((kr_g)^{-1})$, where $J_0$ and $J_1$ are the Bessel-functions of order $0$ and $1$, respectively, and  $x=k^2rr_g(1-\cos\theta)$. To evaluate $\psi_0=e^{\frac{\pi}{2}kr_g}{\Gamma(1-ikr_g)}$, from \cite{Abramovitz-Stegan:1965}, we use the identity  $\Gamma(1-ikr_g)\Gamma(1+ikr_g)={\pi kr_g}/{\sinh \pi kr_g}$, yielding $\psi^2_0={2\pi kr_g}/({1-e^{-2\pi kr_g}}).$
Next, we express the argument $x$  in cylindrical coordinates (\ref{eq:cyl_coord}) as $\sqrt{x}=(\pi{\rho}/{\lambda})\sqrt{{2r_g}/{z}}+{\cal O}(r_g^2,\rho^3).$ For all practical purposes $r_g/r\ll1$; thus, neglecting the corresponding terms and taking into account  $kr_g\gg1$, we  present (\ref{eq:S_rh*3})--(\ref{eq:S_z*3}) in the most relevant form,
{}
\begin{eqnarray}
{\bar S}_z&=&2\pi^2 \frac{r_g}{\lambda}\, J^2_0\Big(2\pi\frac{\rho}{\lambda}\sqrt{\frac{2r_g}{z}}\Big)+{\cal O}(r_g^2, (kr)^{-1}), \quad
 {\bar S}_\rho={\bar S}_\phi={\cal O}(r_g^2, (kr)^{-1}).
\label{eq:S_ph*6}
\end{eqnarray}
As the Poynting vector of a plane EM wave is $ {\bar S}_0=\frac{1}{2}$ \cite{Born-Wolf:1999}, we may  introduce the magnification factor of the SGL as ${\mu}={\bar S}/{\bar S}_0=4\pi^2 ({r_g}/{\lambda})\, J^2_0\big((2\pi\rho/\lambda)\sqrt{{2r_g}/{z}}\big),$ which is valid for small angles $\theta\lesssim\sqrt{2r_g/z}$, i.e.,  in the immediate vicinity of the optical axis \cite{Einstein:1936}. This result represents the SGL's point spread function (PSF), which is a sharply falling and rapidly oscillating function of $\rho$ \cite{Herlt-Stephani:1976,Turyshev-Andersson:2002}. As such, it extends the earlier derivations (e.g., \cite{vonEshleman:1979}) valid only at the optical axis, where $\rho=0$, and provides important details on the structure of the PSF in the interference region of the SGL. 

The wave-optical treatment of the SGL may now be used to consider practical aspects of designing a solar gravitational telescope. 
Equation (\ref{eq:S_ph*6}) suggests that, by naturally focusing light from a distant source, the SGL provides a major brightness amplification (on the focal axis $\mu\sim1.2\times 10^{11}$ at $\lambda=1~\mu$m) and extreme angular resolution ($\lesssim1\times 10^{-10}$ arcsec) in a narrow field of view ($\lesssim3.5$ arcsec) \cite{Turyshev-Andersson:2002}. In fact, starting at 547~AU, the SGL forms a folded caustic, where, in the pencil-sharp region along the optical axis \cite{Herlt-Stephani:1976}, its amplification and angular resolution stay almost unchanged well beyond $10^3$~AU.

An Earth-like planet at 30 parsecs (pc) has an angular diameter of $1.4\times10^{-11}$~rad.
A diffraction-limited telescope comparable in magnifying power to a 1-m telescope placed on the optical axis of the SGL at 750~AU from the Sun would have a diameter of $\sim57$~km. But even this telescope would resolve the disk of the planet only barely. To resolve the planet with $10^3$ pixels across its diameter, one needs a telescope array with a diameter of $\sim4\times10^5$~km ($\sim16R_\oplus$), which is impractical. Building an imaging optical interferometer with a set of such baselines is not feasible. The SGL holds the promise of providing the conditions necessary for a direct megapixel imaging of an exo-Earth.

A modest telescope equipped with a coronagraph could operate at the SGL's focus to provide a direct high-resolution image and spectroscopy of an exoplanet. The image of an exo-Earth is compressed by the SGL into a small region with diameter of $\lesssim5$~km in the immediate vicinity of the focal line.  While all currently envisioned NASA exoplanetary concepts (see {\tt https://exoplanets.nasa.gov/}) aim at getting just a single pixel to study an exoplanet, a mission to the SGL focus opens up the breathtaking possibility of direct imaging (at $10^3\times 10^3$ linear pixels, or $\sim 10$ km in resolution) and spectroscopy of an Earth-like planet up to 30 pc away, enough to see its surface features and signs of habitability.  Such a possibility is truly unique and should be studied in the context of a realistic deep space mission.

In conclusion, a mission to the deep outer regions of the solar system may be able to exploit the remarkable optical properties of the SGL and provide direct megapixel-resolution imaging and spectroscopy of a potentially habitable exoplanet. Although the technical challenges are formidable and have not yet been addressed, the theoretical feasibility and the profound significance of such measurements shall serve as strong motivation to consider the engineering aspects of developing such a mission.

We thank L.D. Friedman, M.V. Sazhin, M. Shao, and V.T. Toth for their interest, support and encouragement during the work and preparation of this manuscript. This work was performed at the Jet Propulsion Laboratory, California Institute of Technology, under a contract with the National Aeronautics and Space Administration.  


\end{document}